# Revisiting the Planet Mass and Stellar Metallicity Relation for Low-Mass Exoplanets Orbiting GKM Class Stars

Jonathan H. Jiang[1], Daniel Zhao[2], Xuan Ji[3], Bohan Xie[4], and Kristen A. Fahy[1]

[1]Jet Propulsion Laboratory, California Institute of Technology, Pasadena, California, USA
[2]Department of Mathematics, Harvard University, Cambridge, Massachusetts, USA
[3]Department of the Geophysical Science, University of Chicago, Chicago Illinois, USA
[4]Department of Physics, Washington University in St. Louis, Missouri, USA

## Abstract

The growing database of exoplanets have shown us the statistical characteristics of various exoplanet populations, providing insight towards their origins. Observational evidence suggests that the process by which gas giants are conceived in the stellar disk may be disparate from that of smaller planets. Using NASA's Exoplanet Archive, we analyzed a correlation between the planet mass and stellar metallicity of low-mass exoplanets ($M_P < 0.13$ $M_J$) orbiting spectral class G, K, and M stars. The correlation suggests an exponential law relationship between the two that is not fully explained by observation biases alone.

## 1  Introduction

Until the first exoplanets were discovered nearly three decades ago, most studies have generally been focused on the properties of gas giants, due to biases of early detection methods like radial velocity towards large planets (Sousa et. al. 2019). In comparison, relatively less is known about the sizeable yet elusive population of small planets. Data on planets smaller than the size of Neptune has recently become more accessible with transit photometry, enabling significant statistical analyses of small planets. Consequently, the field has begun expanding from studying mostly gas giants to exploring sub-Neptune-like planets as well.

Santos et. al. (2004) studied the frequency of planets as a function of the stellar metallicity, while Santos et. al. (2001) also discussed stellar metallicity about the giant planets. These and several subsequent studies have established a strong correlation between the number of giant gas planets in an extrasolar system and the metallicity of their host stars. This is explained by the core accretion model (Pollack et al. 1996) of planetary formation: to form giant planets, enough solid surface density is needed to reach a critical core mass and the stellar metallicity is the threshold for the formation. Based on the database with almost entirely giant planets, this model is consistent with the conclusion that exoplanets are more likely to be found around metal-rich stars (Santos et. al. 2004).

Besides the occurrence-metallicity correlation, a further analysis was discussed in Mordanisi et al. (2012), pointing out the correlation between a planet's initial mass and its host star's metallicity. Given a high metallicity environment, more giant planets are likely to form, but for all ranges of metallicity, the mass distribution is similar. This result is reasonable since, as planetary evolution proceeds, the gaseous envelope composes a majority of the giant planet's mass and not solids. As a result, although the occurrence of the giant planets increases with the metallicity, we cannot easily conclude that the mass also follows the same pattern. In fact, Thorngren et al. (2016) showed that the heavy element enrichment of giant planets relative to their host star's metallicity correlates negatively with planet mass, which suggests that the mass of giant planets does not increase with star metallicity.

As more and more low-mass planets are discovered, Udry et. al. (2007) suggested that the occurrence-metallicity correlation is weaker for sub-Neptune like planets, which is also confirmed in Mordanisi et al. (2012), raising the question: Where is the expected planetary mass increase for small planets orbiting metal-rich stars? To answer this question, the relationship between the mass of low-mass planets and stellar metallicity needs to be examined directly, rather than the frequency of planet occurrence. Recently, Sousa et. al. (2019) presented a metallicity-mass-period diagram of low-mass exoplanets, identifying a potential correlation between the mass of low-mass exoplanets with stellar metallicity. However, no mathematical relationship was established.

## 2  Data

Using the NASA Exoplanet Archive (https://exoplanetarchive.ipac.caltech.edu/) with over 4,300 confirmed exoplanets, we selected a subset of low-mass planets according to three criteria: 1) the projected mass of the planet is < 0.13 Jupiter-mass ($M_J$); 2) its host star is class G, K, or M; and 3) the margin of error for the planet's mass measurement is within 100% of its mass. The range of planets encompassed by these conditions roughly represents the population of sub gas giant exoplanets. Since a major source of uncertainty from radial velocity measurements is noise from stellar activity, and it is shown in Plavchan et al. (2015) that hotter main sequence stars emit greater stellar jitter, the host stars of the sample are limited to low-mass spectral classes to reduce the relative effect of stellar noise on radial velocity (RV) oscillations induced by small exoplanets. In total, 253 planets fitting the conditions were used in this study.

## 3  Results

### 3.1 The Planet Mass - Stellar Metallicity Relation

The focus of this study is to reexamine the relationship between the mass of low-mass planets and stellar metallicity, and to analyze whether the relationship can be explained by observation biases. Figure 1 is the scatter plot for planet mass against stellar metallicity. Two groups of exoplanets emerge: high-mass planets > 0.13 $M_J$ and low-mass planets < 0.13 $M_J$. Note that we also considered 8 different cutoffs from 0.08 $M_J$ to 0.22 $M_J$; however, there are no significant changes with these different values in the two distributions (see Figure A1 in the Appendix).

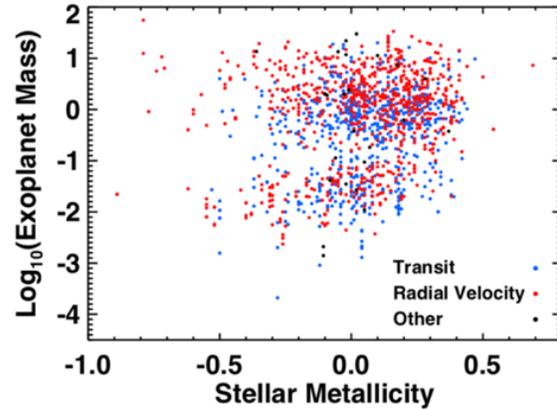

**Figure 1:** Scatter plot of mass against stellar metallicity for all exoplanets. Unit of exoplanet mass is Jupiter-mass. Unit of stellar mass is solar-mass.

Planets detected by transit photometry and radial velocity are well-represented in both the high-mass and low-mass groups, suggesting that the observed group distinction is not due to biases in detection method. While the high-mass planets show no correlation with stellar metallicity, the low-mass planets exhibit a linear relationship with metallicity, as isolated in Figure 2.

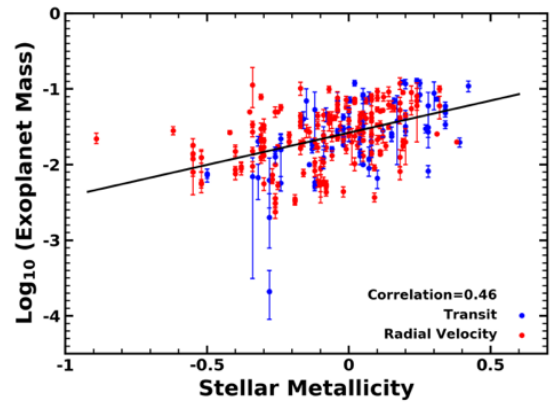

**Figure 2:** Linear regression for mass-metallicity scatter plot of low-mass exoplanets. Unit of exoplanet mass is Jupiter-mass with error margins taken from the NASA exoplanet archive. Unit of stellar metallicity is [$F_e$/H].

Planets detected by transit photometry and radial velocity are well-represented in both the high-mass and low-mass groups, suggesting that the observed group distinction is not due to biases in detection method. While the high-mass planets show no correlation with stellar metallicity, the low-mass planets exhibit a linear relationship with metallicity, as isolated in Figure 2.

To test if this observed trend is significant, a linear regression was performed. The mass-metallicity regression follows the exponential-law equation:

$$\frac{M}{M_J} = a \cdot e^{b \cdot [F_e/H]} \qquad (1)$$

where M is the mass of planet, $M_J$ is Jupiter's mass, $a = 0.28$, $b = 0.85$ with 95% confidence interval (0.65, 1.05), and $[F_e/H]$ denotes the metallicity of the host star. We considered the possibility that the observed trend could be a result of biases inherent to each detection method. A two-dimensional two-sample Kolmogorov-Smirnov test (Fasano, 1987) was performed on both groups with a resulting *p*-value of 0.08, which does not provide sufficient evidence to conclude that the two planet populations are significantly different. Furthermore, planets detected by transit photometry and radial velocity follow the regression line with correlation coefficients of 0.53 and 0.41 respectively over the same metallicity range, suggesting that the relation cannot be entirely attributed to differences in detection methods.

Though some of our sample is discovered by transit photometry, they must undergo follow-up radial velocity observation to obtain the mass measurement (Only 23 planets in our sample are missing the RV amplitude). Another concern is the correlation between stellar mass and stellar metallicity affecting radial velocity measurements. Although metallicity does not directly affect radial velocity measurements, it is correlated with stellar mass (Sousa et. al. 2019), which is a variable in the equation for the semi-amplitude ($K_1$) of a radial velocity oscillation below:

$$K_1 = \left(\frac{2\pi G}{P}\right)^{1/3} \frac{m \sin i}{M^{2/3}} \frac{1}{\sqrt{1-e^2}} \qquad (2)$$

where m, M, P, $i$, and $e$ are the mass of the planet, mass of the star, period, inclination, and eccentricity, respectively (Paddock 1913). Using Kepler's Third Law and assuming a circular orbit and an orbital semimajor axis of 1 Astronomical Unit (AU) gives the detectability relation,

$$M_p \sin i \propto K_1 M_*^{\frac{1}{2}} \qquad (3)$$

between projected planet mass and stellar mass. The inclination (*i*) is randomly distributed so it has no impact on the statistical analysis. From Figure 3, we can see that stellar mass increases approximately linearly with stellar metallicity for this group of low-mass exoplanets, a result that is supported by Leethochawali (2018) and Yates (2012). Assuming a perfect linear correlation between stellar mass and stellar metallicity and performing linear regression to compute $K_1$, Equation (1) follows a steeper exponential trend than Equation (3), a power relation, which is shown in the right panel of Figure 3. Furthermore, the distribution of real RV magnitude of each planet shows no relevance to the exponential trend. Therefore, there is not sufficient evidence to attribute the planet mass stellar metallicity trend to star metallicity-mass correlations.

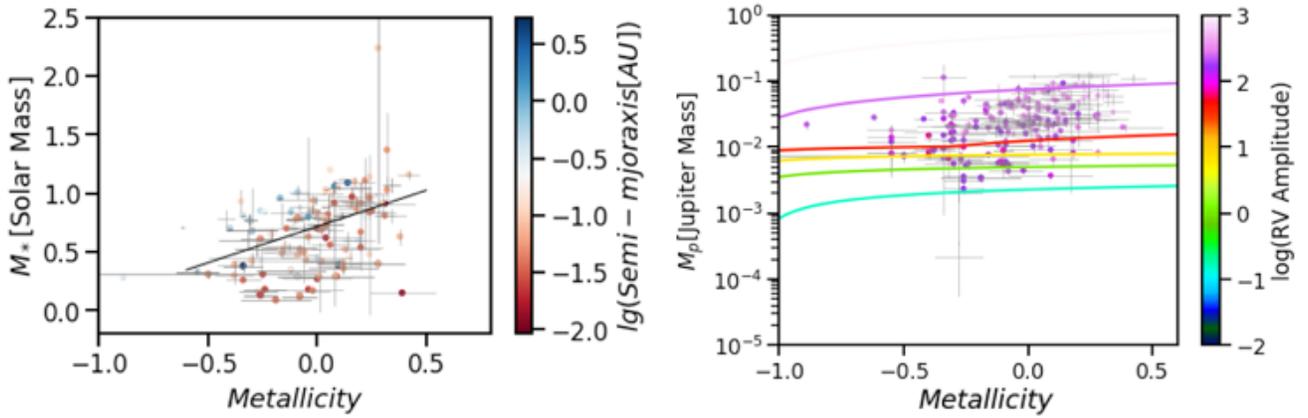

**Figure 3: Left panel:** Stellar mass versus metallicity for low-mass exoplanets. Unit of stellar mass is solar-mass. Unit of stellar metallicity is [Fe/H]. **Right Panel**: Planet mass versus Stellar metallicity. The contour lines mark constant radial velocity amplitudes, which are computed based on the linear relationship of stellar mass and metallicity, while the color of the dots indicate the real RV magnitudes obtained from the NASA Exoplanet Archive website.

### 3.2 The Planet Mass - Stellar Mass Relation

In addition to the mass-metallicity trend, a significant mass-mass relation was also found. Shown in Figure 4, planet mass against stellar mass for the same set of exoplanets was plotted with a linear regression following the equation:

$$\frac{M}{M_J} = a\left(\frac{M_*}{M_\odot}\right)^b \quad (4)$$

with $a = 0.039$ and $b = 0.92$ with 95% confidence interval (0.75, 1.09). Like the mass-metallicity regression, the b slope value for Equation (4) is significantly higher than expected if the trend were due to radial velocity stellar mass scaling. We over-plot the detectability lower bound, as the dotted blue line in Figure 4, using limiting $K_1$ values for 10 m telescopes found in Plavchan et al. (2015), assuming head-on inclination and no eccentricity. The selected population lies well above the bound, suggesting that detectability does not fully explain the trend.

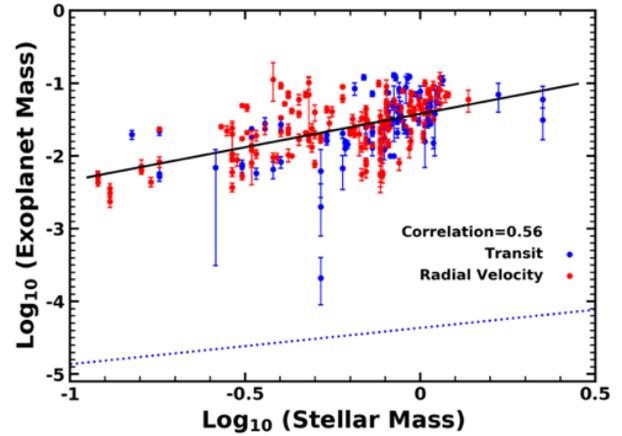

**Figure 4:** Linear regression for mass-mass scatter plot of low-mass exoplanets and lower bound detectability. Unit of exoplanet mass is Jupiter-mass with error margins taken from the NASA exoplanet archive. Unit of stellar mass is solar-mass

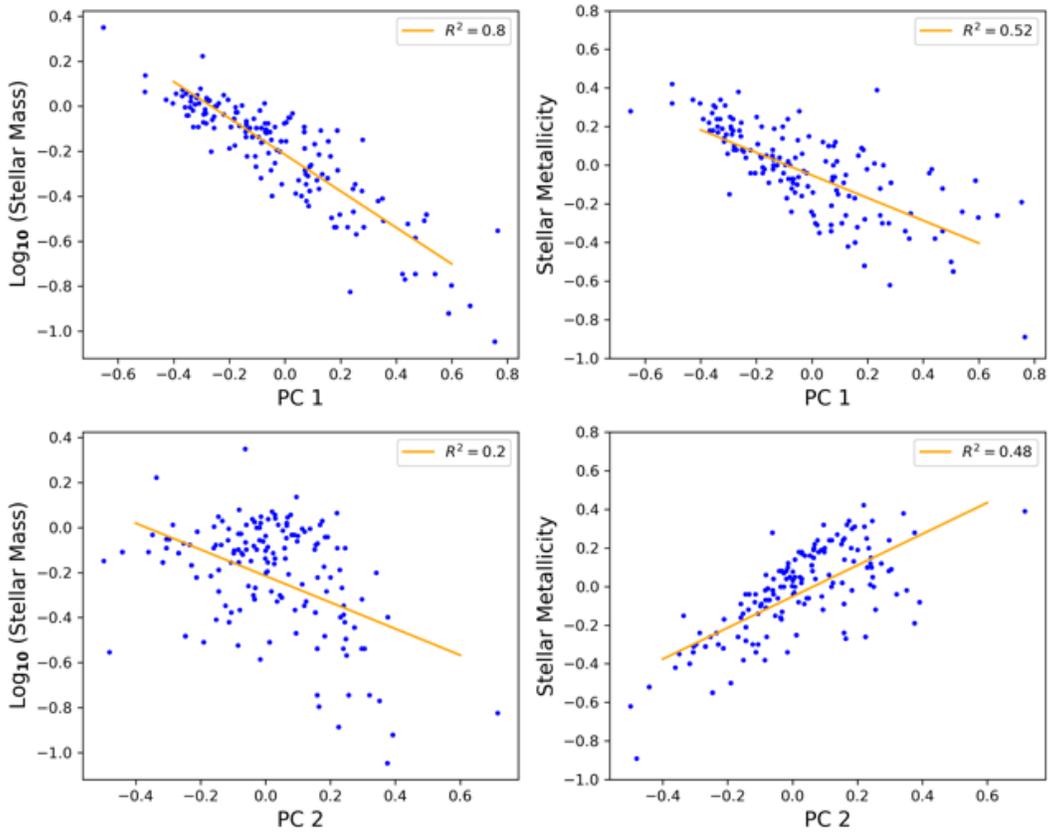

**Figure 5:** Stellar mass and stellar metallicity versus Principal Components. Unit of stellar mass is solar-mass. Unit of stellar metallicity is [Fe/H].

To further investigate the effect of the stellar mass-metallicity relation on this regression, we performed Principal Component (PC) analysis on these two parameters.

PC1 and PC2 capture 68% and 32% of the total variance, respectively. In Figure 5, we plot stellar mass and stellar metallicity against the Principal Components. Although they exhibit somewhat similar distributions along the first Principal axis, a two-dimensional KS test gives a p-value of $5.0\times10^{-10}$, indicating they are significantly different. The difference of distributions is much more apparent along the second Principal axis, as metallicity correlates to PC2 much stronger than does mass, which suggests that a large amount of variance in the data is unaccounted for when considering stellar mass or metallicity alone. It is thus highly unlikely that either the planet mass stellar metallicity trend or the star metallicity-mass are caused by multicollinearity.

### 3.3 Uncertainty Analysis

In our analysis, planet mass is referring to projected mass (M$\sin i$) for those planets without inclination information. It will not influence the linear trend, given that the inclinations ($i$) are randomly distributed, though the absolute value of the index (a and b) would change, if our sample is full of M$\sin i$. However, our sample contains both the projected mass (M$\sin i$) and the accurate mass (M), which could lead to discrepancy between two sublets.

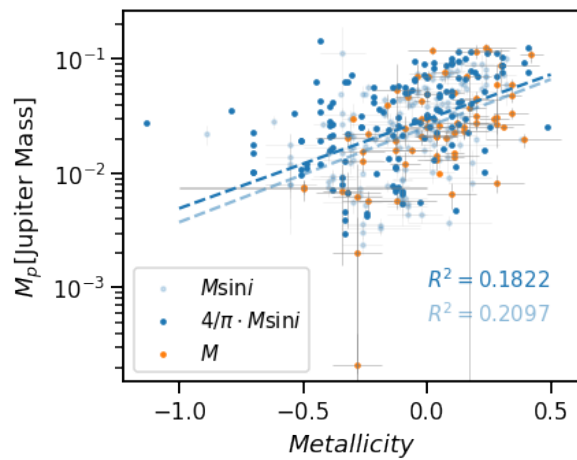

**Figure 6:** The scatter plot of planetary mass versus metallicity. The mass is given in different ways. The orange dots mark those from accurate measurements M; the light blue dots mark the M$\sin i$ measurement value while the dark blue dots mark $4/\pi \cdot$ M$\sin i$. The light and dark blue dashed lines represent the fitting curve from linear regression for the sample of M & M$\sin i$ and M & $4/\pi \cdot$ M$\sin i$, respectively, so do the texts in two colors.

To investigate the impact of this discrepancy on our conclusion, we make up the difference by simply translating M$\sin i$ into its average value of accurate mass ($\langle$M$\rangle$). The distribution function for i is given by $f(i) = \sin i \cdot di$, so the average value of $\sin i$ is equal to $\pi/4$. As a consequence, the average value of mass $\langle$M$\rangle = \pi/4 \cdot$ M$\sin i$. The linear regression for the sample contains accurate mass and projected mass is shown in the light blue in Figure 6, while the linear regression for the sample contains accurate mass and estimated average mass ($\pi/4 \cdot $M$\sin i$) is shown in the dark blue. There is no significant change, indicating the reliability of our analysis.

As M$\sin i$ is the lower limit of the real mass (M), it is sufficient to show the trend of the lower boundary of the planetary mass for a given stellar metallicity. Figure 2 indicates the existence of a linear lower boundary and we can determine the boundary by following the method proposed by Courcol et al. (2016). Firstly, we compute the cumulative distribution function of planetary masses over a succession of metallicity bins which contains 53 data points equally (only the last bin contains 54 data points). Secondly, the 'minimum mass' of the bin is set as different limits (3%,4%,5% and 6%) of this cumulative distribution as shown by the Figure A2 in the Appendix. Thirdly, we conduct linear regression for the minimum mass. For the cutoff limit of 3% which is accepted by Courcol et al. (2016), We determine the lower boundary increasing linearly from 0.0003 $M_J$ to 0.0159 $M_J$, as shown in Figure 7, where the shadow zone indicates a planet desert below the mass lower boundary. To test the sensitivity to the choice of the cutoff limit, we also apply 4%, 5% and 6% as the minimum mass limit and got a similar linear boundary (Figure 7). The lower limit of this boundary ranges from 0.0003$M_J$ to 0.0005$M_J$ and the upper limit ranges from 0.0159 $M_J$ to 0.0199$M_J$. The linear correlation coefficient varies from 1.18 to 1.24 with the relative uncertainty of 5%, which consolidates our conclusion of the lower boundary.

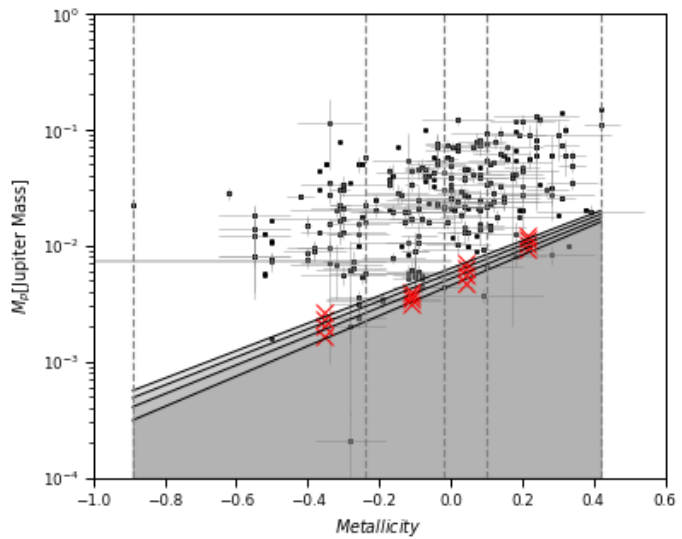

**Figure 7:** The gray dashed lines mark the four bins and the red crosses mark the minimum mass of the bin with different cutoff limits of 3%, 4%, 5% and 6% separately. The lower boundary is computed by linear regression of the minimum mass and the shadow zone indicates the planet desert below the mass lower boundary.

The above analysis suggests the correlation between the planet mass and stellar metallicity is unlikely to be explained by observation biases. The same is true for the planet mass versus stellar mass relation.

## 4   Discussion

An increase in small planet mass associated with higher stellar metallicity is consistent with the core accretion model of planet formation, as dense elements provide matter for solid cores in earlier stages of accretion. Courcol et al. (2016) proposed an upper limit to the mass of Neptune-like planets based on stellar metallicity and orbital period in favor of the in situ formation theory. Our results agree with this finding, but also raise the intriguing possibility that planetary mass may be mathematically tied to stellar metallicity. Currently, the in situ core accretion model does not account for a potential exponential-law relation. The existence of such a trend may suggest that the mass of low-mass planets may be roughly modeled as a function of its host star parameters; the accretion process that leads to the formation of small planets may result in planets whose masses are proportional to the properties of protoplanetary disks – including star metallicity and mass. Due to the relatively low correlation of both trends identified, much more observational data on low-mass planets is needed to confirm this relation.

Another intriguing observation is the apparent difference of properties between gas giant and low-mass planets. While gas giants tend to increase in frequency but not necessarily mass around more metal-rich stars, smaller planets tend to increase in mass; however, there has been no solid evidence that an increase in frequency occurs as well. The disparity between these two populations hints at an undiscovered characteristic of the planetary accretion process. Since terrestrial planets generally have an upper limit of mass before they accrete gaseous envelopes (Lammer et. al. 2014), the lack of a correlation between gas giant mass and stellar metallicity gives reason to believe that the solid cores of gas giants have upper mass limits as well. Otherwise, the total mass of gas giant planets would be expected to increase with stellar metallicity, but this is not the case.

The planet mass stellar mass relation largely agrees with in situ planet formation theories (e.g. Armitage, 2018). Since stellar mass is strongly correlated with the mass of its protoplanetary disk (Andrews 2013), the basic expectation is that total planet mass correlates positively with stellar mass — this condition is fulfilled by the observed trend. If planetary mass can indeed be mathematically modeled using stellar parameters, the power-law relationship identified may provide insight towards the accretion process.

Also noted is that although the solid mass of small planets may decrease slightly on average as stellar mass increases (Mulders 2018), the total mass of small planets increases on average (Figure 4). This might imply that the gaseous envelopes of sub-Neptunian planets increase with stellar mass while the solid core mass decreases, on average. Further studies including the orbital distance of planets may be illuminating, since gas envelope accretion is believed to be only possible at a critical distance from the star (Montmerle 2006).

**Acknowledgement**

This work is supported by the Jet Propulsion Laboratory, California Institute of Technology, under contract with NASA. We also acknowledge the funding support from the NASA Exoplanet Research Program NNH18ZDA001N. We thank Sheldon Zhu for helpful comments and discussions during development of this study.

**Data availability:** All data used in this study are downloaded from the NASA Exoplanet Archive operated by Caltech under contract with NASA (https://exoplanetarchive.ipac.caltech.edu/). The computed data underlying this article are described in the article. For additional questions regarding the data and code sharing, please contact the corresponding author at Jonathan.H.Jiang@jpl.nasa.gov.

Santos, N.C., et al. 2004, A&A, doi:10.1051/0004-6361:20034469.

Sousa, S.G, et al. 2019, *Mon. No. Royal Astron. Soc.*, doi: 10.1093/mnras/stz664.

Thorngren et al. 2016, *The Astrophysical Journal*, doi: 10.3847/0004-637X/831/1/64.

Udry, S, N.C. Santos. 2007, *Ann. Rev. Astron. Astrophys.*, doi:10.1146/annurev.astro.45.051806.110529.

Yates et al. 2012, *Monthly Notices of the Royal Astronomical Society*, doi: 10.1111/j.1365-2966.2012.20595.x
11

# Appendix

In this study, we accept 0.13 Jupiter mass as the cutoff, which is a subjective threshold estimated from the visual gap apparent in Figure 1, taking previous literature on low-mass exoplanets as reference (e.g. Sousa et al. 2019, which has a cutoff at 0.094 Jupiter mass). We also considered 8 different cutoffs from 0.08 $M_J$ to 0.22 $M_J$, however it can be seen from the results, $R^2$ are not very sensitive to the threshold regardless. It is suggested that the correlation holds even with many different cuts.

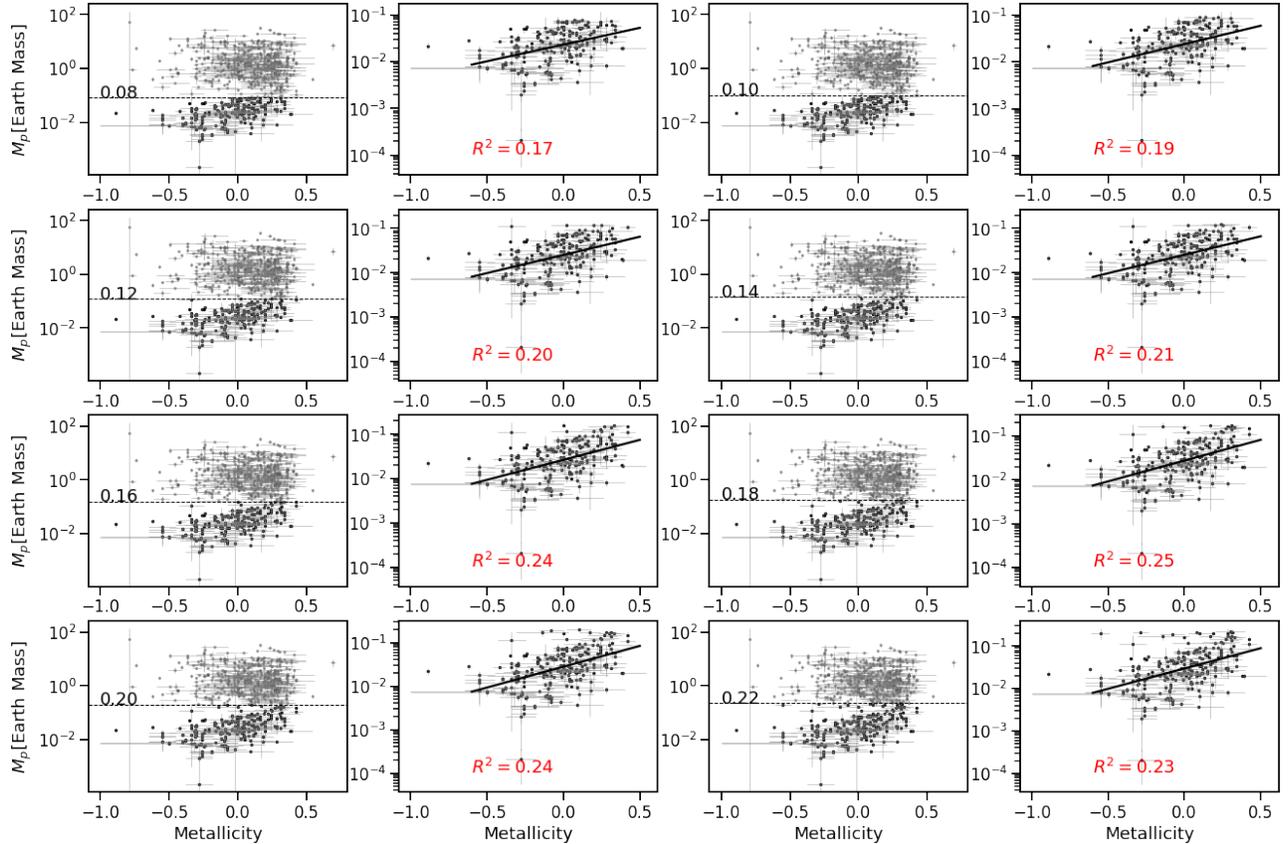

**Figure A1:** The first and third columns show scatter plots for planetary mass against the metallicity of planets satisfying our criterion (2) and (3). The gray dots represent the high-mass planets, of which the masses are above different cutoffs as shown by the gray horizontal lines, while the black dots represent the low-mass planets of which the masses are lower than the cutoffs. The linear regressions for those low-mass planets are shown to the left and the fitting scores ($R^2$) are shown on the figure.

To determine the lower boundary of the planetary mass for a given boundary, we divide the sample into 4 subsets regarding a succession of metallicity bins. The first 3 bins contain 53 data points and the last bin contains 54. For each bin, we compute the cumulative distribution function of planetary masses and the 'minimum mass' of the bin is set as different (3%,4%,5%,6%) limits of this cumulative distribution as shown by Figure A2. It shows that the results are not sensitive to the value of the cut-limit, which validify our conclusion of the existence of the lower limit.

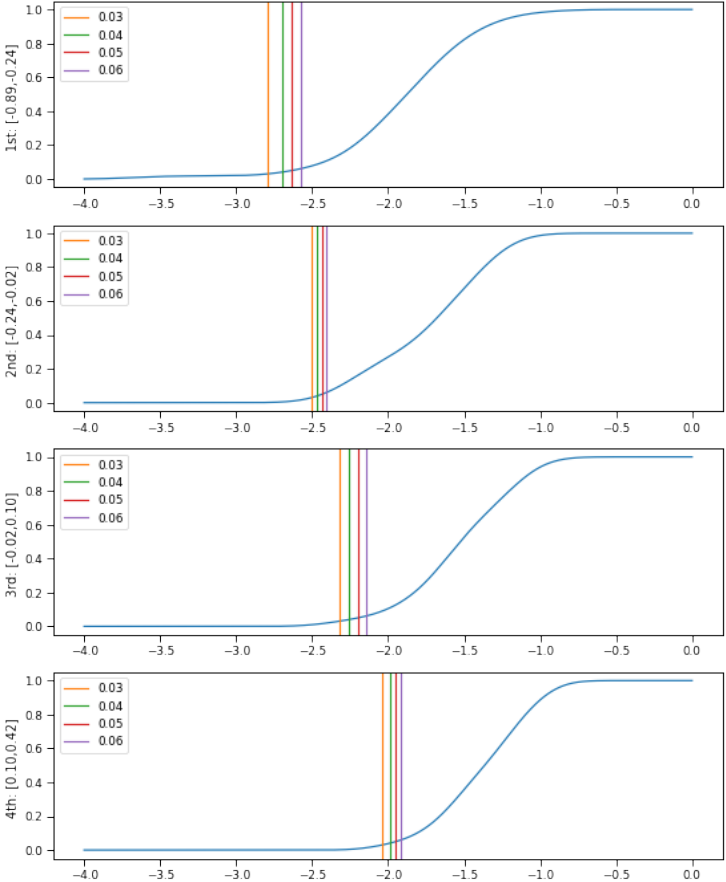

**Figure A2:** The cumulative distribution function of planetary masses over a succession of metallicity bins from the bottom to the top, and each bin contains 53 data points, except that the last bin has 54 data points. The ranges of bins are shown on the left y-label. The vertical lines mark the "minimum mass" of the bin is set as the 3%,4%,5%,6% limit of this cumulative distribution as shown in the legend.